\pdfoutput=1

\documentclass[preprint,amsmath,amssymb]{aastex63}

\usepackage{multirow}
\usepackage{amsmath}
\usepackage{rotating}
\usepackage{float}
\usepackage{xcolor}
\usepackage[normalem]{ulem}
\usepackage{blindtext}
\graphicspath{{./}{temp_figures/}}

\begin{document}

\title{Astrophysical Evidence of Wakefield Acceleration in Galactic and Extragalactic Jets via Gamma Rays and UHECRs}

\correspondingauthor{Toshikazu Ebisuzaki}
\email{ebisu@postman.riken.jp}

\author{Gregory B. Huxtable}
\affiliation{UC Irvine, Physics and Astronomy}

\author{Noor Eltawil}
\affiliation{UC Irvine, Physics and Astronomy}

\author{Wei-Xiang Feng}
\affiliation{UC Riverside, Physics and Astronomy}

\author{Wenhao Wang}
\affiliation{UC Irvine, Physics and Astronomy}

\author{Gabriel Player}
\author{Toshiki Tajima}
\affiliation{UC Irvine, Physics and Astronomy}

\author{Toshikazu Ebisuzaki}
\affiliation{Computational Astrophysics Laboratory, RIKEN}











\begin{abstract}

We present six case studies from a broad mass range ($1 - 10^9$ $M_\odot$) of astrophysical objects, each of which exhibit signs of jets and emit intense high energy gamma rays ($>10$ GeV).  Many of these objects also emit spatially identifiable ultra high energy cosmic rays (UHECRs).  
In all cases it is found that wakefield acceleration (WFA) explains both the global properties and details.
For blazars, we also explain the temporal structure of these signals, which includes neutrinos, and the correlations in their ``bursts" and anti-correlation in flux and index.
Blazars ($\sim 10^9$ $M_\odot$), 
radio galaxies ($\sim 10^8\, M_{\odot}$), Seyfert galaxies ($\sim 10^6 \,M_{\odot}$), starburst galaxies ($\sim 10^{3}\, M_{\odot}$),
down to microquasars ($1 \sim 10$ $M_\odot$) interestingly exhibit the same physics since the nature of the accretion and acceleration is independent of the mass, aside from maximum values.
It is possible to accelerate electrons to energies much greater than $10$ GeV, and protons beyond $10^{20}$ eV with WFA.
We compare observational values with theoretical ones to illustrate they are in good agreement.
This mechanism is also accompanied by related emissions, such as high-energy pin-pointed neutrinos, time varying radio, optical, and X-ray emissions, opening an opportunity to characterize these astrophysical objects via multi-messenger approaches.

\end{abstract}



\section{Introduction} \label{sec:intro}
We note that a wide class of astrophysical objects ranging from blazars, radio galaxies, Seyfert galaxies, starburst galaxies, and microquasars emit intense high energy gamma rays (10 GeV), often in spatially (localized) and temporally identifiable fashions. We opt to study typical objects from each respectable class of astrophysical categories as case studies to examine their emission mechanism and accompanying signals. These signals include gamma-ray emissions, ultra high energy cosmic rays (UHECRs), optical emission, radio, X-ray emissions, and possibly neutrino emission. The selected objects are all spatially localized. Some also exhibit temporal structure in their signals, characteristic of wakefield acceleration (hereafter referred to as WFA). The emissions often are simultaneous, such as the  coincidental temporal signals of high energy gamma rays and neutrinos.  Inspired by such spatially pinpointed emissions and temporal coincidences in signal type, we quantitatively compare these observational features with a theory that can give rise to them, i.e. the WFA \citep{EbisuzakiandTajima2014a,EbisuzakiandTajima2014b,TYE2020}. It is argued that the compact central astrophysical objects (active galactic nuclei, or active stellar binaries) can accompany accretion disk and jets \citep{Shibata1986, PlAstr}. Under these circumstances, in spite of the disparate central masses amongst the selected objects, similar physical sequences and observational features may be expected. The structure and dynamics, we study in theory and find in observations, lead to unique and characteristic features spatially, temporally, and in energy. Also, the various emission signals (such as high energy gamma rays, UHECRs, high energy neutrinos, optical, radio, X-rays, etc) observed, and predicted from theory are relatively close in value. Other, less structured acceleration mechanisms, such as Fermi acceleration \citep{fermi54} and turbulent heating/acceleration are compared in contrast.

The high energy phenomena associated with accreting blackholes has been observed with gamma-ray emission. First, The Fermi Gamma-ray Space Telescope, formerly GLAST (Gamma-ray Large Area Space Telescope) launched in 2008 \citep{Michelsonetal2010} observed accreting blackholes such as active galactic nuclei (AGN) and binary blackholes in the GeV-100 GeV region \citep{Abdoetal2010,Ackermannetal2011,Ackermannetal2012}. In addition, air Cherenkov telescopes, such as MAGIC \citep{Djannati-Atai2009}, H.E.S.S. \citep{Djannati-Atai2009}, or VERITAS  \citep{Ragan2012} and water Cherenkov detector such as High Altitude Water Cherenkov (HAWC) observatory \citep{DeYoung2012} observed accreting blackholes in TeV and multi TeV gamma rays.

Cosmic rays vary from modest to extremely high energies ($10^{20}$ eV and possibly beyond).  The conventional theory by Fermi mechanism \citep{fermi54} has been successful in explaining the universal spectrum index of approximately 2 \citep{KoteraReview} for UHECRs.  However, beyond $10^{19}$ eV, protons begin to radiate their energies very quickly via synchrotron radiation if they are bent by magnetic fields (as Fermi mechanism assumes) or other collisions \citep{Jackson}. Also the Fermi mechanism, based on its stochastic acceleration by the galactic magnetic fields, renders that cosmic rays are coming from all directions in roughly equal amount. On the other hand, observations have detected cosmic rays with energies beyond $10^{19}$ eV coming from localized origins \citep{Abraham2008, Heetal2016, DiMatteo2019}.

\begin{figure}[H]
    \centering
    \includegraphics[scale=0.5]{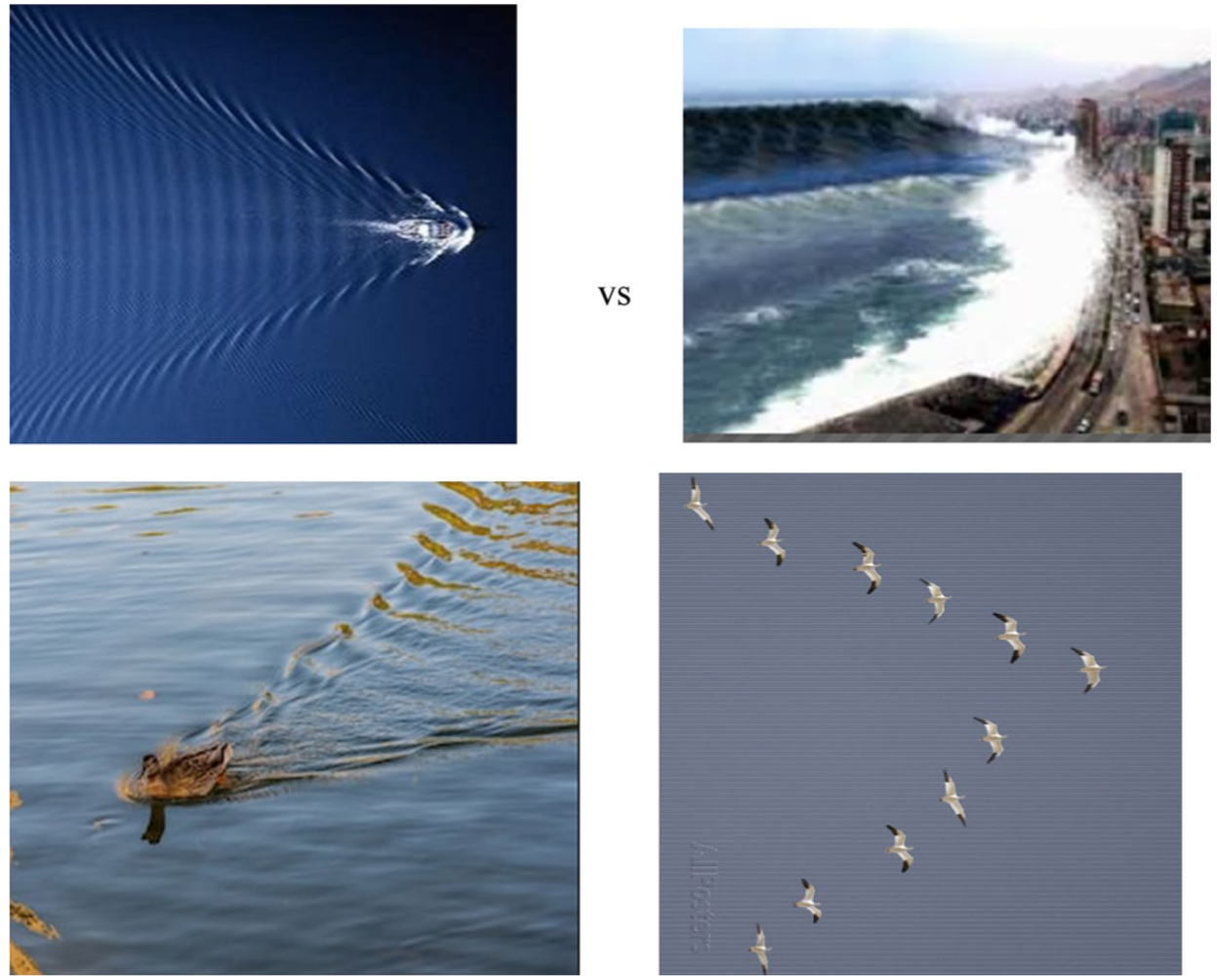}
    \caption{\label{fig:epsart1} Three examples of wakefield in nature. (Top left) a boat travelling on a smooth lake creates a wave (wake) behind it with phase velocity equal to the speed of the boat.   (Bottom left) a duck, similar to the boat, creates a wave behind it with phase velocity equal to the duck's.  (Bottom right) birds flying in a chevron pattern conserve energy by riding in the bow shock created by the lead bird, which is reinforced by the successive birds.
     (Top right) On the other hand, a tsunami waves gains amplitude and turbulence takes over, showing the importance of the high phase velocity principle \citep{TYE2020}.
    }
\end{figure}

\begin{figure}[ht!]
   \centering
   \begin{tabular}{cc}
      \begin{minipage}[t]{0.6\textwidth}\centering
      \includegraphics[width=\textwidth]{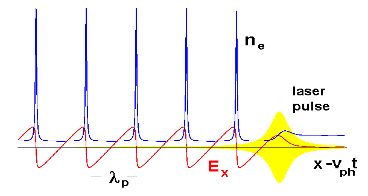}
      \end{minipage}
      
   \end{tabular}
\caption{\label{fig:epsart2} Wakefield simulation depicting E-field in red, electron density in blue, and the instigating laser pulse in yellow. The clear coherency due to the high phase velocity (in fact relativistic) wakefields is demonstrated. (after \cite{TYE2020}).}
\end{figure}

\begin{figure}[t]
   \centering
   \begin{tabular}{cc}
      \begin{minipage}[t]{0.75\textwidth}\centering
      \includegraphics[width=\textwidth]{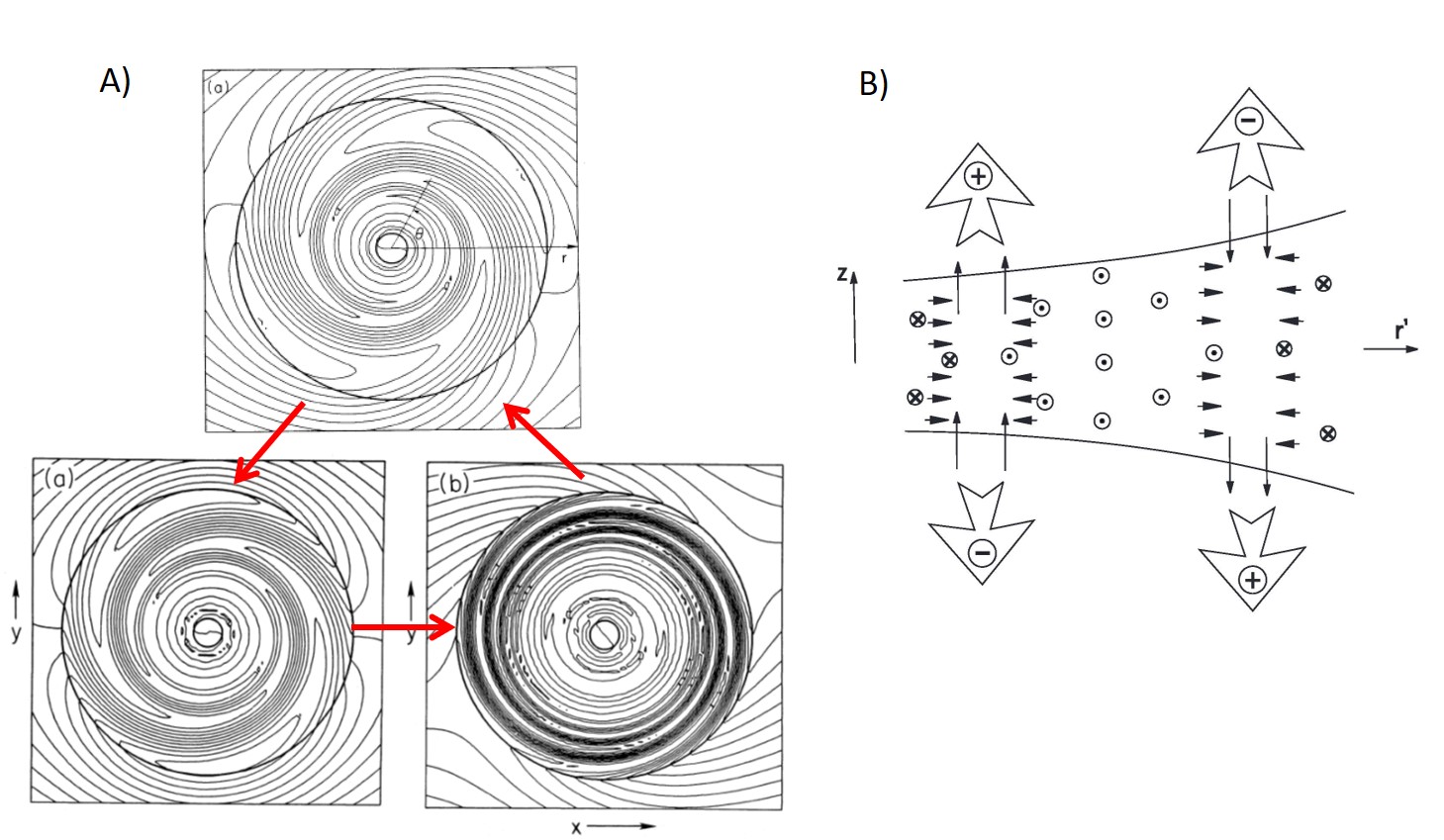}
      \end{minipage}
      
   \end{tabular}
\caption{\label{fig:MRI} The physical consequences of the magneto-rotational instability. A) Magnetic field is amplified by the winding motion due to the differential rotation of the accretion disk. When the field strength becomes too strong to be maintained by the accretion disk, it springs back to the less magnetized state, emitting a burst of Alfvénic disturbances \citep{Gilden, Haswell}. B) Magneto-Rotational Instability (MRI) produces an electro-static field, which accelerates charged particles and heats up the disk halo. They eventually produce low energy photons in keV-MeV energy region \citep{Haswell}.
}
\end{figure}
The emerging new mechanism of wakefield acceleration of \cite{TDPRL} has quite a different theoretical construct (Figures \ref{fig:epsart1}, \ref{fig:epsart2}).  It is based principally on a single astrophysical object such as an AGN and its accretion disk and associated jets. The accretion disk instability such as the magnetorotational instability (MRI) \citep{MRI, Gilden} can introduce the rapid increase of the present magnetic fields in the disk, which triggers episodic disruption of the accretion disk and subsequent disturbances at the feet of the jets \citep{Mizuta2018}.  See figure \ref{fig:MRI}. This disturbance may be considered as the trigger of intense electromagnetic (originally Alfvénic shock) pulses in the jets.
According to \cite{TYE2020} we evaluate the basic physical parameters for the range of astrophysical objects using WFA.  

\begin{figure}[H]
   \centering
   \begin{tabular}{cc}
      \begin{minipage}[t]{0.8\textwidth}\centering
      \includegraphics[width=\textwidth]{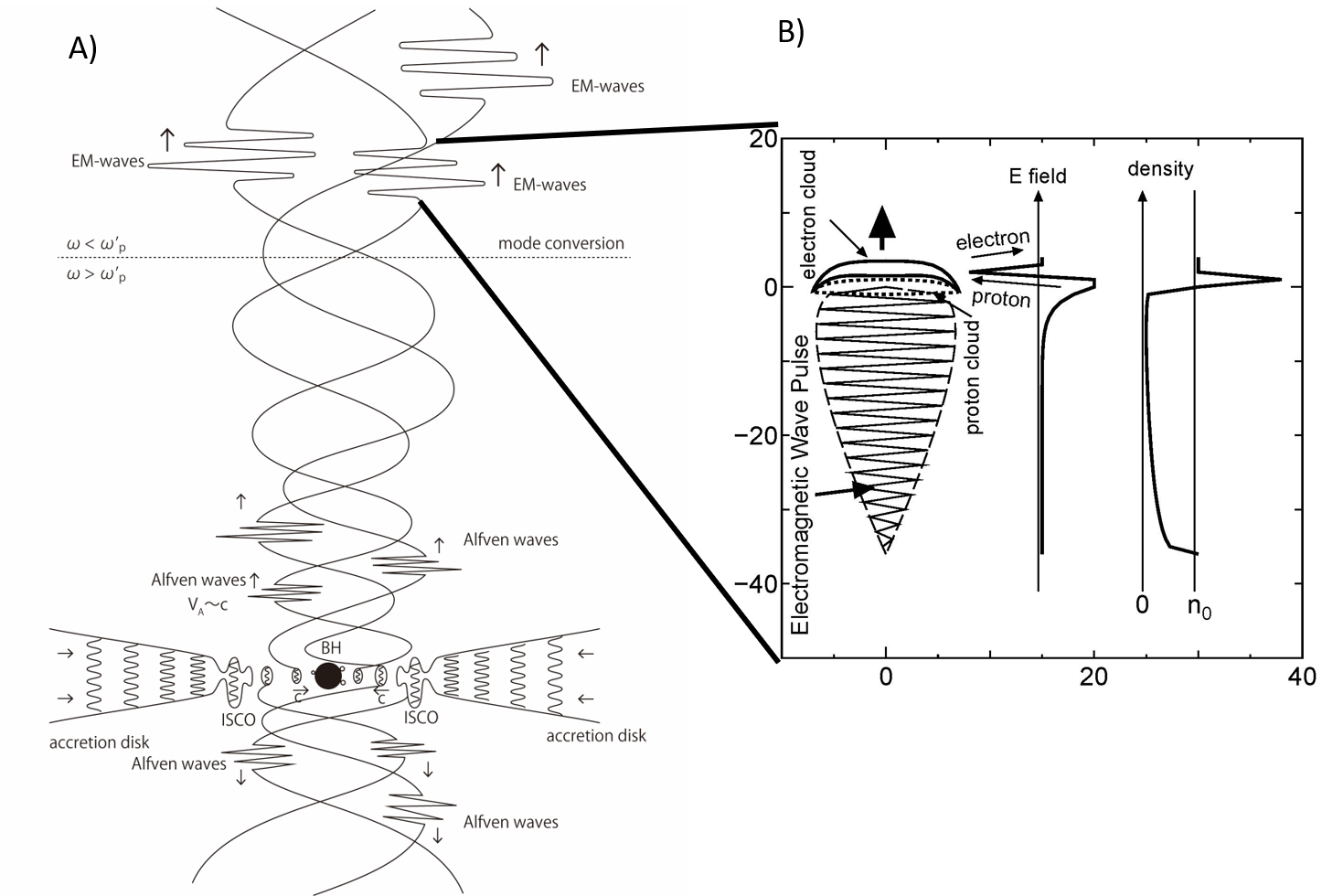}
      \end{minipage}
      
   \end{tabular}
\caption{\label{fig:bowwake} WFA theory for the accreting black hole-jet system of an AGN \citep{EbisuzakiandTajima2014a,EbisuzakiandTajima2014b,TYE2020}. A) The cross section of an AGN, its accretion disk and jets. Alfvén waves generated upon large mass accretions at the base of the jet propagate along the field line of the jet. They eventually mode-convert into intense electromagnetic waves. B) The structure of the bow wake. An electron cloud is formed at the front (top) of the wave pulse and a proton cloud follows. The resultant electric field accelerates protons in the back side and electrons in the front side of the bow wake. Since this acceleration structure moves at a velocity close to the light velocity, the charged particles with the same velocity (light velocity) in the field are accelerated for a long time.
}
\end{figure}

In the WFA theory, developed by T. Ebisuzaki and T. Tajima \citep{EbisuzakiandTajima2014a,EbisuzakiandTajima2014b,TYE2020}, input parameters are $m$ and $\dot{m}$. The former, $m$, the BH mass, normalised by the solar mass ($M_{\odot}=2.0\times 10^{33}\, \rm g$), can be estimated by observations such as stellar dynamics, QPO/recurrence period, and Eddington limit. The later is the accretion rate $\dot{m}$, normalised by critical accretion rate, $\dot{M_{\rm c}}=\frac{2\pi c R_0}{3\epsilon \kappa_{\rm T}}$. Here, $\kappa_{\rm T}$ is the Thomson scattering opacity and $\epsilon=0.06$ is the radiation efficiency of the disk. The non-dimensional accretion rate $\dot{m}$ can be calculated using the radiation luminosity $L_{\rm rad}$: 
\begin{equation}
\label{L_rad}
\begin{split}
L_{\rm rad}=\frac{4\pi cG M_{\odot}}{\kappa_{\rm T}}\dot{m}m,
\end{split}
\end{equation} 
which is proportional to the product of $\dot{m}$ and $m$. The radiation luminosity $L_{\rm rad}$ can be estimated by emission lines from the nucleus for the case of active galactic nuclei (blazars, radio galaxies, and Seyfert galaxies) or by the X-ray luminosity for the case of less massive black holes ($<1000 M_{\odot}$: intermediate and stellar mass blackholes). Since X-rays are believed to be emitted directly from the accretion disk by thermal mechanisms (not from the jets by non-thermal mechanisms) in the less massive blackholes, their anisotropies are minimal compared with those of gamma rays. 

When $m$ and $\dot{m}$, or equivalently $L_{\rm rad}$ are given, WFA theory predicts the rise time $2\pi/\omega$ of the burst, the episodic recurrence time $1/\nu$, and the acceleration time $D_3/c$, as follows:
\begin{equation}
\label{2pi/omega}
\begin{split}
2\pi/\omega=\frac{R_0}{6\epsilon c}\alpha^{1/2}\dot{m}m,
\end{split}
\end{equation}

\begin{equation}
\label{1/nu}
\begin{split}
1/\nu=\frac{\sqrt{6}R_0}{c}\alpha^{-1/2}m,
\end{split}
\end{equation}

\begin{equation}
\label{D3/c}
\begin{split}
D_3/c=\frac{1}{36}\left(\frac{e^2R_0^4}{4\pi^3c^5m_{\rm e}^2 \epsilon^5\kappa_{\rm T}}\right)^{1/3}\alpha^{5/6}\dot{m}^{5/3}m^{4/3},
\end{split}
\end{equation}
where $R_0=6GM_{\odot}/c^2=9.0\times10^5 \, \rm{cm}$ is the radius of the innermost stable orbit of a one solar mass black hole, $c$ the light velocity, $\alpha$ the ``alpha'' disk parameter, $e$ and $m_{\rm e}$ are the electron charge and mass, respectively.

The protons are accelerated in the back side of the bow wake (see figure \ref{fig:bowwake}B). 
The maximum proton energy (UHECRs) is given as:
\begin{equation}
\label{W_max}
\begin{split}
W_{\rm max}=\frac{1}{9}\bigg(\frac{e^4c^2R_0^2}{2m_e\epsilon^4\kappa_{\rm T}^2}\bigg)^{1/3}z\Gamma\alpha^{2/3}\dot{m}^{4/3}m^{2/3},
\end{split}
\end{equation}
where $\Gamma$ is the bulk Lorentz factor of the jet. 
The luminosity of UHECR is calculated as: 
\begin{equation}
\label{L_UHECR}
\begin{split}
L_{\rm UHECR}=\frac{\zeta \sigma \alpha^{1/2}}{6\epsilon}L_{\rm rad},
\end{split}
\end{equation}
where $\sigma$ is the the energy efficiency of the charged-particle acceleration, including the conversion of Alfven wave into electromagnetic waves, and
\begin{equation}
\label{zeta}
\begin{split}
\zeta=\frac{\ln (W_{\rm max}/W_0)}{\ln (W_{\rm max}/W_{\rm min})}.
\end{split}
\end{equation}
Here, $W_0=0.57\times 10^{20} \rm eV$.

On the other hand, the electrons are accelerated in the front side of the bow wake (figure \ref{fig:bowwake}B), simultaneously with protons in WFA. The accelerated electrons emit high energy gamma rays with the energies ranging GeV-PeV due to the collision with the magnetic perturbations. 

The luminosity of gamma photons is given as:
\begin{equation}
\label{L_gamma}
\begin{split}
L_{\gamma}=\frac{\sigma \alpha^{1/2}}{6\epsilon}L_{\rm rad},
\end{split}
\end{equation}
though the energies and spectrum shape are different depending on the situation of the jets.

For the reader's convenience, we summarize the scaling laws represented by the equations above in Table \ref{tab:ScalingLaws}.
 
 The flux, $F_{\rm UHECR}$, at the Earth is calculated as:
\begin{equation}
\label{F_UHECR}
\begin{split}
F_{\rm UHECR}=6.7\times 10^{-1}  \left[\frac{\rm UHECRs}{100\,\rm km^2  \, yr}\right]
\left(\frac{d}{3.6\, \rm Mpc}\right)^{-2}
\left(\frac{L_{\rm rad}}{10^{42}\,\rm erg\, s^{-1}} \right),
\end{split}
\end{equation}
for the case of isotropic radiation with $\rm ln(W_{\rm max}/W_{\rm min})=30$, $\alpha=0.1$ and $\sigma=0.1$.

The cosmic ray protons, accelerated by the wakefield, may collide with another proton in the plume of decelerated material in the jet or interstellar gas in the object to produce pions, which decay into gamma rays, electrons, and neutrinos. The neutrino number flux, $F_\nu$, of neutrinos arriving at Earth can be obtained  from:
\begin{equation}
\label{eq:F_nu}
    W_\nu^2F_\nu=\frac{f_{pp}\sigma \alpha^{1/2} L_{\rm rad}}{6\Omega  d^2\epsilon \ln \left(W_{\rm max}/W_{\rm min}\right)},
\end{equation}
where $f_{\rm pp}$ is the collision probability of protons, $\Omega$ is the solid angle of the emission ($\Omega=4\pi$ for the isotropic case) and we assume $W_\nu=0.05W_{\rm p}$.

We consider the likely scenario of electromagnetic pulses produced in the jets near the innermost part of the accretion disk accelerating charged particles (protons, ions, electrons) to very high energies ($10^{20}$ eV for  protons and ions and $10^{12-15}$ eV for electrons) by electromagnetic wave-particle interaction via wakefields \citep{TYE2020}. The episodic, eruptive accretion in the disk by the magneto-rotational instability gives rise to the strong electro-magnetic pulses, which act as the drivers of the collective acceleration of the pondermotive force. This pondermotive force drives the wakes. The accelerated hadrons (protons and nuclei) are released to the intergalactic space and some eventually reach, and decay in, Earth's atmosphere as  UHECRs. Some of them collide with protons in the interstellar medium to produce secondary particles, such as neutrinos and gamma-rays. The high-energy electrons, on the other hand, emit photons as a result of collisions with electromagnetic perturbances to produce various non-thermal emissions (radio, IR, visible, UV, and gamma rays).

We apply WFA theory to six candidates: Blazars, such as TXS 0506+056; radio galaxy, Centaurus A; Seyfert galaxy, NGC1068; starburst galaxies, M82 and NGC 0253; and microquasar, SS 433. We survey and scrutinize general commonalities as well as specific characteristics of their various signals, including cosmic rays and very high energy (VHE) gamma rays.  These six astrophysical objects have vastly varying central masses from $10^9$ solar mass down to only several solar mass. Nonetheless, their polar jets exhibit common phenomena of intense ion acceleration as well as electron acceleration simultaneously, fundamentally in a linear fashion with a pulsed operation as mentioned in episodic motion associated with the central object’s accretion disk variations. Thus, the UHECRs are pointed as well as accompanied by gamma-ray emission (due to the electron acceleration) with a specific separation in the arrival time of the signals.Neutrino arrival may coincidental, but delayed in time as well. In other words, though the mass scales are vastly varied, the underlying mechanisms are remarkably common.  Thus, the burst periods and rise and fall times may differ among cases, but the mass dependence and qualitative features are curiously common.   These are what we wish to investigate in detail for the previously mentioned six astrophysical objects with various central masses.  The emerging picture is a surprisingly unified, integrated physical mechanism of wakefield acceleration.

\begin{figure}[H]
   \centering
   \begin{tabular}{cc}
      \begin{minipage}[t]{0.6\textwidth}\centering
      \includegraphics[width=\textwidth]{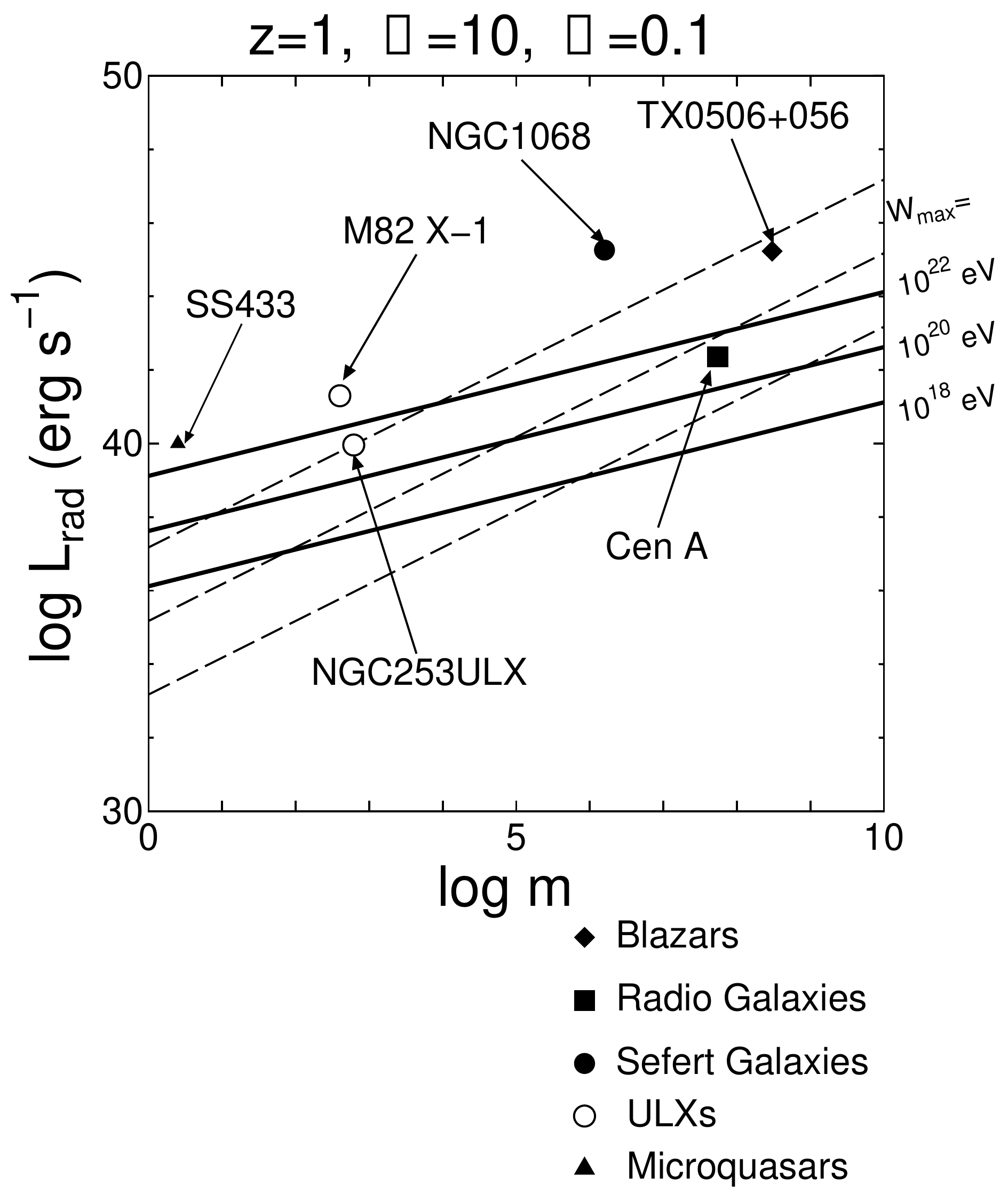}
      \end{minipage}
      
   \end{tabular}
\caption{\label{fig:Wmax} The maximum energies of protons, $W_{\rm max}$, are predicted using the mass of the central normalized black holes, $m$, on the right abscissa, and total radiation luminosity, $L_{\rm rad}$, by the wakefield acceleration theory \citep{TYE2020} on the left abscissa. Various astronomical objects, such as microquasars ($m=1\sim 10$), ultra luminous X-ray sources (ULXs; $m=100\sim 10000$) in starburst galaxies, central blackholes of Seyfert galaxies $m\sim 10^6$ and radio galaxies/blazars ($m=10^7\sim 10^{10}$) can accelerate UHECRs with energy above $10^{20}\,\rm eV$.  The individual points (or boxes) are plotted based on their known (or estimated) central mass according to our theory. Here, we assume the charge is $z=1$, the bulk Lorentz factor of the flow of the jets $\Gamma=10$, and $\alpha=0.1$ the ``$\alpha$" parameter of the disk. From top to bottom, the three dashed lines represents $\dot{m}=10^{-1}$, $10^{-3}$, and $10^{-5}$.}
\end{figure}

Before we close this section, let us mention a separate, but related important issue of the recent simultaneous observation of the gravitational waves \citep{MisnerThorneandWheeler1973} and gamma rays from the collision of two neutron stars \citep{LIGO2017,Abbottetal2017}. When LIGO observed the gravitational wave (GW) arrival \citep{Abbottetal2017}. It was suggested \citep{Takahashietal2000} that a collision of two neutron stars can yield not only the emission of violent phenomena such as GWs, but also gives rise to the formation of an accretion disk and its jets.  It follows that an eruption at the base of the jets, such as a massive accretion, produces wakefield acceleration of electrons (and thus gamma photons), following the emission of a GW. Thus, we see that the gamma emission is an important indicator of the underlying physical process of the electron acceleration (by WFA) and alerts us to the importance of the multi-messenger astrophysics approach.

In the proceeding sections we describe our study of six astrophysical objects that have been observed to have broad  range of central BH mass (from $\sim 10^9 \ M_\odot$ to $\sim 10 \ M_\odot$) in descending order.  They are: blazars in Sec. \ref{sec:Blazar}), Centaurus A (radio galaxy) in Sec. \ref{sec:leve3}, NGC 1068 (Seyfert galaxy) in Sec. \ref{sec:NGC1068}, M82, (starburst galaxy) in Sec. \ref{sec:M82}, NGC 0253 (starburst galaxy) in Sec. \ref{sec:NGC 0253}, and SS 433 (microquasar near our galactic center) in Sec. \ref{sec:leve6}.   Localized UHECRs or neutrinos are observed in all cases (except SS 433), in addition to highly luminous UHE gamma rays. In Sec. \ref{sec:summary} we summarize the comparison of our research of these astrophysical objects with their observations, and their derived acceleration processes.

\begin{table}[H]
\caption{\label{tab:ScalingLaws}
Time Scales, Maximum Energy, and Luminosities predicted by \citet{TYE2020}}
\begin{ruledtabular}
\begin{tabular}{cccc}
 Quantities&Scaling Laws&Units&Equation Numbers\\ \hline
 $2\pi/\omega$&$8.2\times10^{-5}\alpha^{-1/2}\dot{m}m$&s&\ref{2pi/omega}\\
 $1/\nu$&$7.3\times10^{-5}\alpha^{-1/2}m$&s&\ref{1/nu}\\
 $D_3/c$&$1.7\times10^2\alpha^{5/6}\dot{m}^{5/3}m^{4/3}$&s&\ref{D3/c} \\ \hline
$W_{\rm max}$&$3.2\times10^{-31}z\Gamma\alpha^{2/3}m^{-2/3}L_{\rm rad}^{4/3}$&eV&\ref{W_max}\\
 \hline
 $L_{\rm rad}$&$1.5\times10^{38}\dot{m}m$&erg $\rm s^{-1}$&\ref{L_rad}\\
 $L_{\rm \gamma}$&$2.78\sigma\alpha^{1/2}L_{\rm rad}$&erg $\rm s^{-1}$&\ref{L_gamma}\\
 $L_{\rm UHECR}$&$2.78\sigma\zeta\alpha^{1/2}L_{\rm rad}$&erg $\rm s^{-1}$&\ref{F_UHECR}\\
\end{tabular}
\end{ruledtabular}
\end{table}

\section{\label{sec:Blazar} Blazar:  TXS 0506+056}

TXS 0506+056 in constellation Orion is a  blazar, a quasar with a relativistic jet pointing directly towards Earth, with a redshift of $0.3365 \pm 0.0010$ \citep{Paianoetal2018}, which corresponds to about 1.75 Gpc from Earth.
TXS 0506+056 was first cataloged as a radio source in 1983 \citep{Lawrenceetal1983}, and then confirmed a blazar \citep{Massaroetal2009}. Gamma rays were detected by the EGRET and Fermi-LAT missions \citep{LambandMacomb1997,Halpernetal2003,Abdoetal2010_2}. In addition, radio observations have shown apparent superluminal motion in the jet \citep{Richardsetal2011}.

Furthermore, on 22 September 2017, the cubic-kilometer IceCube Neutrino Observatory detected a high-energy neutrino emission from a direction consistent with this flaring gamma-ray blazar TXS 0506+056. The most probable energy for the observed neutrino is around $190$ TeV with a 90\% confidence level (CL) lower limit of $183$ TeV, depending only weakly on the assumed astrophysical energy spectrum. Such observation may imply the existence of extremely high energy protons or nuclei with tens of PeV generated in the jet of the Blazar. The clear emission direction with high energy particles may suggest a different acceleration mechanism for the ultrahigh energy cosmic rays other than Fermi's stochastic acceleration \citep{fermi54}.

\begin{figure}[H]
   \centering
   \begin{tabular}{cc}
      \begin{minipage}[t]{0.9\textwidth}\centering
      \includegraphics[width=\textwidth]{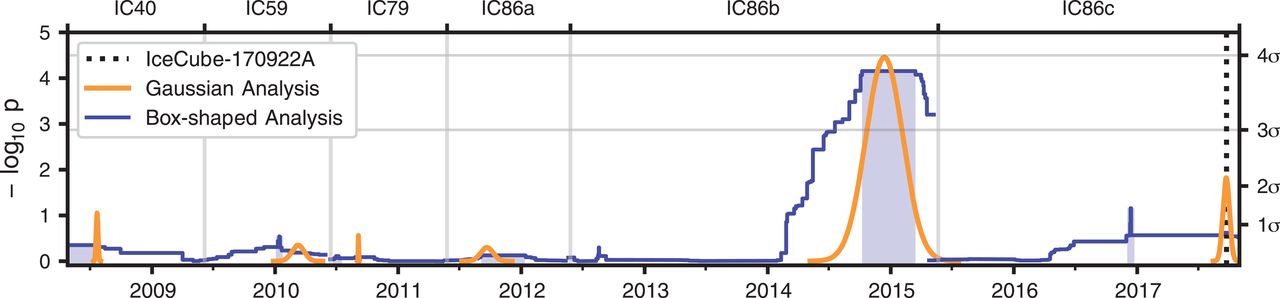}
      \end{minipage}
      
   \end{tabular}
   \caption{Periodic bursts are shown in the neutrino data from \cite{IceCube2018b}}
   \label{fig_neutrino_data}
\end{figure}

During $\pm 2$ weeks of the neutrino observation, a peak flux of gamma ray emission around $5.3\times 10^{-7}\, \rm cm^{-2}\, s^{-1}$ is also reported by Fermi-LAT, with an energy range between $0.1\sim 100$ GeV \citep{IceCube2018a}. The associated isotropic luminosity during the period reaches as high as $1.2\times 10^{47} \, \rm erg\, s^{-1}$\citep{IceCube2018b}. Analysing the data prior to the event, a long-term isotropic gamma ray luminosity between $0.1$ GeV and $100$ GeV is derived with an averaged value of $0.28\times 10^{47}$ erg/s over 9.5 years of Fermi-LAT observations of TXS 0506+056 \citep{IceCube2018b}. According to their study, the Gaussian-shaped time profile of the neutrino emission shows a periodic burst pattern depicted as in Fig.\ref{fig_neutrino_data}, which is similar to MRI instability relaxation depicted in \cite{Canacetal2020}.

%

The high energy phenomena including neutrino and photons encouraged us to try WFA as an alternative explanation for the neutrino emission event. According to the WFA theory, the relativistic ponderomotive acceleration in the jet can boost particles to an energy over ZeV \citep{EbisuzakiandTajima2014a,EbisuzakiandTajima2014b,TYE2020}. Also, because the particles are accelerated linearly, the corresponding signal detected can be highly localized around the location of the blazar.  In the rest of the section, we will estimate several physical parameters using the wakefield theory and compare them with their analogous observational value.

Mass estimation is difficult in general for blazars. \citet{padovani19}, however, estimated the central black hole mass to be $3\times 10^8\, M_{\odot}$ using the relations of black holes mass and R-band bulge magnitude $M(R)\sim -2.9$ \citep{Paianoetal2017,Paianoetal2018}, assuming the host galaxy to be a giant elliptical. The bolometric luminosity ($\simeq L_{\rm rad}$) of $1.7 \times 10^{45}\,\rm erg\, s^{-1}$ is taken into account of the over estimate due to jet-induced component using the bolometric luminosity derived from the the OII and OIII lines ($L_{\rm OII}=9\times 10^{45}\,\rm erg\, s^{-1}$ and $L_{\rm OIII}=3\times 10^{45}\,\rm erg\, s^{-1}$). Substituting $m=3\times 10^8$ and $L_{\rm rad}=1.7\times 10^{45}\,\rm erg\, s^{-1}$ into equations \ref{W_max}, \ref{2pi/omega}, \ref{1/nu}, and \ref{L_gamma}, we derived $W_{\rm max}=3.1\times 10^{24}\rm eV$, $2\pi/\omega=3.0\times 10^3\, \rm s$, $1/\nu=6.9\times 10^4\, \rm s$, and $L_{\gamma}=1.5\times 10^{44} \,\rm erg\, s^{-1}$, as shown in Table \ref{tab:bigtable}.

Although the maximum proton energy, $W_{\rm max}$ for protons is well above $10^{20}$  eV, UHECR protons propagate only 100 Mpc and can not reach to the Earth because of the GZK mechanism \citep{Greisen1966,ZatsepinandKuzmin1966} in which protons undergo inelastic collisions with  photons in the cosmic microwave background loose energy. 

According to the MAGIC observation \citep{Ansoldi2018}, very high energy (VHE) gamma rays, above 90 GeV, from TXS 0506+056 varied, increasing by a factor of 6 within a day. We may set $2\pi/\omega=4.8\times 10^4\, \rm s$ or shorter as an e-raising time. There are two periods (2017 October 3-4, and 2017 October 31) of enhanced gamma-ray emission. We may set $1/\nu$ as $2.4\times 10^6\,\rm s$ or shorter, taking into account the incomplete observation in TeV gamma rays.

The theoretical gamma-ray luminosity is calculated as $L_{\gamma}=1.5\times 10^{44}$ erg s$^{-1}$ (see Eq. \ref{L_gamma}), which is much less than the observed isotropic gamma-ray luminosity $1.2\times 10^{47} \, \rm erg\, s^{-1}$ observed by FERMI-LAT \citep{IceCube2018b}. This is probably due to the concentration of the radiation being axially aligned with the jets.

If the actual emission is strongly beamed, such isotropic assumption will overestimate the luminosity by integrating the whole sphere. If we assume a diameter of $10^\circ$ for the beam, the corrected estimation of the Fermi-LAT luminosity will be 2 orders lower than their reported value. That is, $L_\gamma^{(10^\circ)} = 4.6 \times10^{44}$ erg/s, which is consistent with the theoretical luminosity.
 
Here, $W_{\rm \nu}$, and $W_{\rm p}$ are the energies of the neutrino and the proton, respectively. Substituting $f_{\rm pp}=1.0$, $L_{\rm rad}=1.7\times10^{45} \, \rm erg\, s^{-1}$, d=1.8 Gpc, $\sigma=0.1$, $\alpha=0.1$, $\epsilon=0.06$, $\Omega/4\pi=4.8\times 10^{-2}$ for $\theta=10^{\circ}$, and $\ln (W_{\rm max}/W_{\rm min})=30$ into equation \ref{eq:F_nu}, we obtain $F_\nu=8.1\times 10^{-16} \, \rm TeV^{-1}\, cm^{-2}\, s^{-1}$ at 100 TeV. It is consistent with the observation by IceCube of $F_\nu=1.6\times 10^{-15}\, \rm TeV^{-1}\, cm^{-2}\, s^{-1}$ at 100 TeV \citep{IceCube2018b}.

The WFA theory can be a good candidate in explaining the simultaneous arrival of the high energy gamma ray and neutrino flux in the blazar direction, because in the theory the particles are linearly accelerated and hence follow a clear direction. The theory gives an underestimated gamma ray luminosity compared to the observation data. Such discrepancy may be explained by the localization of the beam-like emission in contrast to the isotropic emission which is commonly believed to be true (see Appendix). The periodicity of neutrino bursts in past data may be qualitatively explained by the burst due to the MRI instability, the same burst that accelerates hadrons to UHECRs and produces VHE gamma rays. Although it can be difficult to detect the highest energy particles at present, it is promising that  future studies may find more evidence to support our theory. Provided with more accurate physical data for the accretion disk of the blazar, we may also be able to refine our calculation 

\section{\label{sec:leve3}Radio Galaxy: Centaurus A}

Centaurus A (also known as NGC 5128), in constellation Centaurus, is a radio galaxy with kpc size jets, and at a distance of 3.4 Mpc from the Earth \citep{CenA}. Centaurus A hosts an AGN (active galactic nuclei) believed to be a supermassive blackhole with the mass of $5.5\times 10^7 M_{\odot}$ \citep{Neumayeretal2010}.  The jets extending from the blackhole are the result of the magneto-rotational instability (MRI) \citep{MRI,Gilden} and expelled accreted matter dragging field lines, and allow for extremely fast evolution of the galaxy as matter is accelerated by WFA near the speed of light \citep{TYE2020}.

The mass of the Cen A is well defined by stellar dynamics \citep{Neumayeretal2010} as $m=5.5\times10^7$. The $L_{\rm rad}$ is also determined to be $L_{\rm rad}=2.3\times 10^{42}$ erg s$^{-1}$ from X-ray observations \citep{Jourdain1993}.
Substituting $m=5.5\times 10^7$ and $L_{\rm rad}=2.3\times 10^{42}\,\rm erg\, s^{-1}$ into equations \ref{W_max}, \ref{2pi/omega}, \ref{1/nu}, and \ref{L_gamma}, we derived $W_{\rm max}=1.4\times 10^{21}\rm eV$, $2\pi/\omega=4.2\, \rm s$, $1/\nu=1.3\times 10^4\, \rm s$, and $L_{\gamma}=2.0\times 10^{41} \,\rm erg\, s^{-1}$, as shown in table \ref{tab:bigtable}.

\begin{figure}[H]
    \centering
    \includegraphics[scale=0.8]{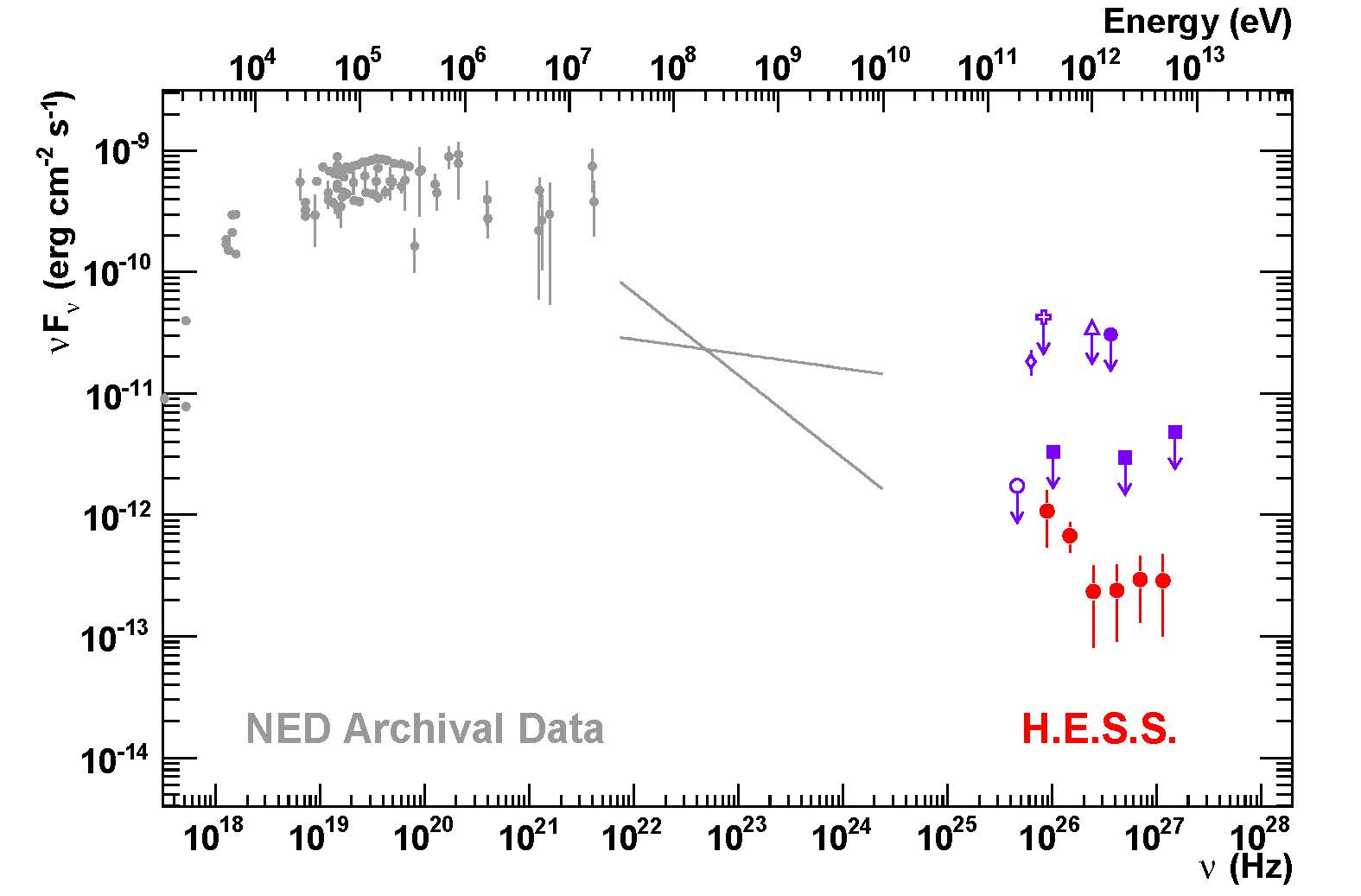}  
    \caption{\label{fig:CenAspectra} Observed gamma-ray spectrum of Cen A distinctly shows evidence of gamma ray emission at 10 TeV \citep{Aharonian2009}.}
\end{figure}

\begin{figure}[H]
\epsscale{1.0}
    \centering
    \includegraphics[scale=0.8]{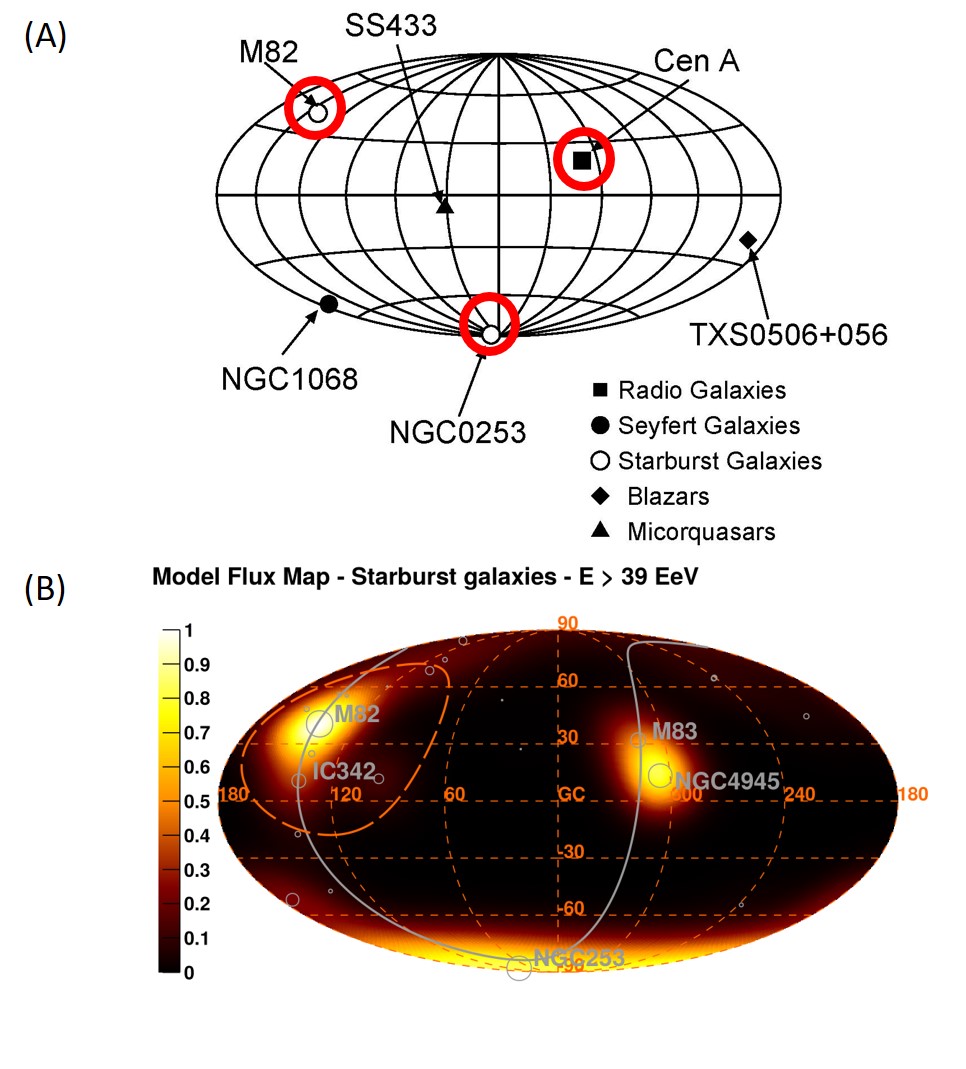}
    \caption{\label{fig:skymap} Prominent candidates of pin-pointed emission of UHECRs. A) Skymap of the possible UHECR sources.Wakefield acceleration theory predict M82, NGC0253, Cen A, and NGC4945 are the promising UHECR sources (red circles). B) Three hot spots in the observed skymap of UHECRs $>6\times10^{19}$ eV \citep{Aab2018} consistent with the theory prediction.}
\end{figure}

The WFA theory predicts that Cen A is capable of accelerating protons to  energies above $10^{20}$ eV (see Table \ref{tab:bigtable}). If we assume isotropic emission of UHECR, the theoretical UHECR flux turns to be $0.60\, \rm /km^2\,/yr$.
However, \cite{Aab2018} observed  UHECRs as energetic as $10^{20}$ eV coming from Cen A (figure \ref{fig:skymap}). The observed UHECR flux is about $0.016\, \rm /km^2\,/yr$, which is consistent within  a factor of five with the theoretical prediction (see table \ref{tab:bigtable}).
The episodic recurrence time predicted by WFA theory is consistent with the observation of \cite{Fukazawaetal2011,Rothschildetal2011}, who showed $50\%$ time variability in the time scale of $10-20$ ks ($1/\nu\sim 1.5\times 10^4$). 

The theoretical gamma-ray luminosity for Cen A is calculated as $L_{\gamma}=2.0\times 10^{41}\, \rm erg\, s^{-1} $. On the other hand,
Cen A has been known to emit gamma rays in the range of TeV and greater for decades now, and the The H.E.S.S. telescope determined the gamma-ray flux of $0.45\pm 0.07 \times 10^{-13}\,\rm ph\, cm^{-2}\, s^{-1} TeV^{-1}$ at 1 TeV (Figure \ref{fig:CenAspectra}). The corresponding gamma-ray luminosity calculated as $1.1\times 10^{38} \,\rm erg\, s^{-1}$ \citep{Abdalla2018}.  
The factor of $\sim 1000$ difference between the theoretical and observational  value for $L_\gamma$ is most likely due to the fact that the axis of the jets are at a large angle to our line of sight, and gamma-ray emissions are strongly beamed in the axially direction of the jets.  

%

\begin{table}[H]
\caption{Comparison of observed parameters (shaded) and theoretical parameters (without shade).  Theoretical  and observational  values for a range of astrophysical objects, including a blazar (BL), a radio galaxy (RG), a Seyfert galaxy (SyG), starburst galaxies (SBG) and a micro-quasar (MQ).
}
\label{tab:bigtable}
\begin{ruledtabular}

\begin{tabular}{c|cccccc}
parameter & TX 0506+056 & Cen A & NGC1068 & M82 & NGC 0253 & SS 433 \\
 
\hline

type & BL & RG & SyG & SBG & SBG & MQ \\
\hline\hline
\colorbox{lightgray}{$\log d$ (pc)} & 9.24 & 6.53 &7.15 & 6.56 & 6.54 & 3.54 \\

\colorbox{lightgray}{$\log M_{\rm BH}(M_{\odot})$} & 8.48 & 7.74 & 6.20& 2.60 & 2.79 & 0.40\\

\colorbox{lightgray}{$\log L_{\rm rad}(\rm erg\,s^{-1})$} & 45.23 & 42.36 & 45.26& 41.30 & 39.96 & 40.00\\
\hline
$\log W_{\rm max}\,(\rm eV)$ & 24.49 &21.16&26.05 &23.17 & 21.26 & 22.91\\
$\log L_{\rm UHECR}\,( \rm erg\, s^{-1})$ & 43.17 & 40.31 & 43.20&39.24 & 37.90 & 37.94\\
$F_{\rm UHECR}(/100\,\rm km^2/yr)$ & - & 0.69 & -&0.052 &0.0026 & -\\
\colorbox{lightgray}{$F_{\rm UHECR}(/100\,\rm km^2/yr)$} & - & 0.016 &-& 0.040& 0.013 & -\\
\hline
$\log 2\pi/\omega \,\rm (s)$ & 3.47 & 0.62 &3.50& -0.46 & -1.81 & -1.76\\
\colorbox{lightgray}{$\log 2\pi/\omega\,\rm (s)$} & $<4.68$ & - & -&- & - & -\\
\hline
$\log 1/\nu \,\rm (s)$ & 4.84 & 4.10 &2.56& -1.04 & -0.85 & -3.24\\
\colorbox{lightgray}{$\log 1/\nu,\rm (s)$} & $<6.38$ & 4.18 & -&-0.70 & $<2.0$ & $<1.0$\\
\hline
$\log L_{\gamma}\,( \rm erg\, s^{-1})$ & 44.17 & 41.31 & 44.20&40.24 & 38.90 & 38.94\\
\colorbox{lightgray}{$\log L_{\gamma}\,( \rm erg\, s^{-1})$} & 47.08 & 38.04 & 45.53&40.18 & 39.78 & 37.57\\
\end{tabular}
\end{ruledtabular}
\end{table}

\section{\label{sec:NGC1068} Seyfert Galaxy NGC 1068}

NGC1068 is a Seyfert galaxy that has a bright nucleus with a central blackhole of $1.6\times 10^6\, M_{\odot}$ \citep{Gouldingetal2010}. It is also undergoing an intense starburst process. It is located at a distance of 14 Mpc \citep{tully1988} in the constellation Cetus. The bolometric luminosity of the nucleus of $1.8\times10^{45}\,\rm erg\, s^{-1}$ was obtained from OIV line \citep{Gouldingetal2010}. 
Substituting $m=1.6\times 10^6 $, $L_{\rm rad}=1.8\times 10^{45}\,\rm erg\, s^{-1}$, and $\alpha=0.1$ into equations \ref{W_max}, \ref{2pi/omega}, \ref{1/nu}, and \ref{L_gamma}, $W_{\rm max}=1.1\times 10^{26}\, \rm eV$, $2\pi/\omega=3.2\times 10^{3}\,\rm s$, $1/\nu=3.63\times 10^2\,\rm s$, $L_{\gamma}=1.6\times 10^{44}\,\rm erg\, s^{-1}$, respectively (Table \ref{tab:bigtable}).

As $W_{\rm max}>10^{20}\, \rm eV$, the nucleus of NGC 1068 has an ability to  efficiently accelerate UHECRs. In fact; isotropic UHECR is as high as $\sim 30 \,\rm UHECRs/100\, km^{2}\, /yr$ (Table \ref{tab:bigtable}), though the spot size would be too large (70 degree or more) because the distance to the Earth is three times larger than M82, Cen A, and NGC 0253 ($\sim 3\,\rm Mpc$) due to the intergalactic magnetic field \citep{Globusetal2008}.

There is no significant luminosity change in the intrinsic luminosity from  the accretion disk of the nucleus of NGC 1068. Although \citet{Zainoetal2020} reported the time variability in the time scale of 1-6 months, the detailed spectral analysis revealed that the variability is not due to the change in the intrinsic accretion rate but due to the change in obscuring Compton thick cloud ($N_H>10^{25}\,\rm  cm^{-2}$) \citep{Mattetal2004}, which surrounds the nucleus. This view is consistent with the infrared and optical observations \citep{TaranovaandShenavrin2006,HoenigandKishimoto2011}.

\citet{Ackermannetal2012} observed isotropic gamma-ray luminosity of $1.5\times 10^{41}\,\rm erg\, s^{-1}$. It is consistent with the theoretical prediction above and Table \ref{tab:bigtable}, although it is just by chance, since the gamma-ray flux is likely to be beamed. GeV gamma rays are generally believed to be from supernova remnants in the galaxy as the result of intense starburst activity.
On the other hand, \citet{Acciarietal2019} set an upper limit of NGC 1068 in gama-ray above 200 GeV at $5.1\times 10^{-13}\,\rm erg \,cm^{-2}\, s^{-1}$. It corresponds to the isotropic luminosity of $3.5\times 10^{45}\,\rm erg\, s^{-1}$, which is one order of magnitude large compared with the theoretical prediction.

The IceCube collaboration reported the positive detection of neutrinos at 1 TeV of $3 \times 10^{-13}\,\rm TeV^{-1}\, cm^{-2}\, s^{-1}$ from NGC $1068$.
Substituting $m=1.6\times 10^6$, $L_{\rm rad}=1.8\times 10^{45}\,\rm erg\, s^{-1}$, $\Omega=4\pi$, and $\alpha=0.1$ into equation \ref{eq:F_nu}, we obtain the theoretical isotropic flux as: $2.2\times 10^{-11}\,\rm TeV^{-1}\, cm^{-2}\, s^{-1}$.
In other words, WFA theory can explain IcuCube observation, if 1\% of the neutrinos emitted from the jets travel towards Earth; the neutrinos from the jets are most likely to be strongly beamed.

\section{\label{sec:M82} Starburst Galaxy: M82}

M82 is a starburst galaxy of the distance of 3.6 Mpc \citep{Freedmanetal1994} in the constellation Ursa Major. 
The starburst activity takes place in a relatively small central region, radius of $\sim200$ pc \citep{Volk1996} from the dynamic center of the galaxy.

Ultra luminous X-ray (ULX) sources of luminosity $\gtrsim10^{40}\, \rm erg\, s^{-1}$ inside M82 have been observed (see \citet{Xu2015} \& the references therein). Among them, M82 X-1 is the brightest ULX in M82, located about 200 pc away from the dynamic center of the galaxy \citep{Matsumotoetal1999,Tsuruetal2004,Patrunoetal2006,Dewanganetal2006,Feng2010}.  There has been a lot of discussions in the recent 20 years regarding the mass of M82 X-1, which has converged to the intermediate mass range, in other words $10^2-10^3\ M_\odot$. We adopt the mass of $400\, M_\odot$ by \cite{Pasham2014}, who used QPO frequency to fit a mass value. 
Substituting $m=4\times 10^2$ and $L_{\rm rad}=2\times 10^{41}\, \rm erg\, s^{-1}$, and $\alpha=0.1$ into equations \ref{W_max}, \ref{2pi/omega}, \ref{1/nu}, and \ref{L_gamma}, we derived $W_{\rm max}=1.5\times10^{23}\,\rm eV$, $2\pi/\omega=3.5\times 10^{-1}\, \rm s$, $1/\nu=9.1\times 10^{-2}\, \rm s$, and $L_{\gamma}=1.7\times 10^{40}\, \rm erg \, s^{-1}$, as shown in Table \ref{tab:bigtable}.

The WFA theory predicts that M82 X-1 has an ability to accelerate protons and nucleus to UHECR with the energy above $10^{20} \rm eV$, in spite of its less massive blackhole ($\sim 400\, M_{\odot}$; Table \ref{tab:bigtable}), unlike the Fermi acceleration theory.  In fact; the Telescope Array (TA) team suggested that there is a hot spot in the northern sky of arrival direction of the UHECRs above 57 EeV (\citet{Abbasietal2014}, see also  Figure \ref{fig:skymap})in close proximity to M82.

\citet{Heetal2016} divided the events belonging to the northern hot spot into two by energy, and found that there was a systematic deviation between them. Assuming that this is due to the deflection by the magnetic field \citep{Globusetal2008}, the position of the true source was estimated. While the estimated position, though extended to 10 degrees, included several high-energy celestial objects such as M82 and Mrk 180, only M82 was located within the GZK-horizon (~100 Mpc) that the UHECRs could reach. 
The Telescope Array (TA) team detected 72 cosmic rays of 57 EeV in 5 years. Among them, 19 events are within the hot spot \citep{Abbasietal2014}, while 4.5 events were expected from uniform arrival. Since the effective area of TA is $700\,\rm km^2$, the observed excess flux in the hot spot direction is about $0.040\,\rm UHECRs/100\,km^{2}/yr$, which is consistent with the expected isotropic flux from equation \ref{L_UHECR} ($\sim 0.052\,\rm UHECRs/100\,km^{2}/yr$, as shown in Table \ref{tab:bigtable}); the direction of UHECRs may be randomized, due to the strong magnetic field inside M82.

The QPO period of M82 X-1 is observed in X-ray band to be $0.2$ s \citep{Pasham2014}. As we have carried out in Sec. \ref{sec:Blazar}-\ref{sec:leve3}, the theoretical recurrence time is  $1/\nu=9.23\times 10^{-2}$ s (Table \ref{tab:bigtable}). This predicted value is well consistent with the QPO period within a factor of 2.

In the WFA theory, electrons are also accelerated in the similar way with protons (see Figure \ref{fig:bowwake}B). The high energy electrons, accelerated by the wakefield in the direction of the jet, emit gamma rays by synchrotron process with the interaction with magnetic perturbations in the jets. This gamma-ray luminosity of M82 X-1 can be also calculated as $1.7\times 10^{40}\, \rm erg \, s^{-1}$ (Table \ref{tab:bigtable}). The energy spectrum is likely to be expressed by a single power law from  GeV to 100 TeV, with an constant index, which is close to 2 in the strongest acceleration case, depending on the magnitudes of the acceleration field and the magnetic field in the jets \citep{Canacetal2020}. Note that this gamma-ray emission is expected to  be strongly concentrated in the direction of the jets.

On the other hand, a bright and isolated gamma-ray excess, consistent with the location of the position of M82 of 100 MeV to $700$ GeV gamma-rays that are isotropic in luminosity of $1.5\times 10^{40}$ erg/s with FERMI-LAT \citep{Ackermannetal2012}).  This is consistent with the theoretical prediction, though it might be just by chance, taking into account of the non-isotropic nature of WFA theory. 

Several flux points of $\gamma$-ray energy of $>700$ GeV 
were observed in M82 \citep{Abdoetal2010}. The fitted power-law spectrum suggests that a single physical emission mechanism, such as WFA theory, dominates from GeV to TeV energies for M82, though a popular explanation is those from numerous number of supernova remnants in the nucleus region of starburst galaxy M82. 

Time variabilities have not yet been reported from Fermi-LAT observations \citep{Ackermannetal2012}. If it is the case in future observations, that will be an evidence that gamma-ray emission comes from a compact object, such as M82 X-1, not from the extended sources, like a supernova remnant.  

\section{\label{sec:NGC 0253}Starburst Galaxy: NGC 0253}

NGC 0253 is a nearly edge-on starburst galaxy located at the distance of $3.5 \pm 0.2$ Mpc from the Earth \citep{Rekolaetal2005} in the constellation Sculptor. \citet{Aab2018} reanalyzed the data of arrival direction observed by Pierre Auger Observatory (PAO) and found that a significant (4$\sigma$ level) enhancement in the arrival direction map of UHECRs above 39 EeV with the search radius of 12.9 degree toward nearby starburst galaxies, NGC 0253 (Figure \ref{fig:skymap}). The result is consistent with the data of Telescope Array team, though statistically marginal \citep{Aab2018, Attallah2018}.

\citet{Gutierrez2020} proposed two candidates for the source of UHECRs, one is TH2 \citep{Turner1985} and the other is NGC253 X-1. Although TH2 was presumed the brightest radio source nearly coincident to the center of the galaxy, the recent observation by ALMA revealed that the
position of TH2 exactly coincides to one of the knots in the central region of NGC 0253, which are most likely HII regions excited by young compact star clusters. The mass of the clusters are estimated as $~10^{6} M_{\odot}$ and are not
likely to have any blackholes, since there are no X-ray emissions. Although one may still assume a hidden non-accreting black hole in the cluster, any blackhole without accretion cannot emit any energy.
\citet{Gutierrez2020} assumed a strong magnetic field of $10^4$ G around the blackhole to produce jet luminosity through the Blandford-Znajek effect \citep{BlandfordandZnajek1977}. However, this magnetic field will decay rapidly if no accretion on the blackhole. The luminosity in equation 3 of \citet{Gutierrez2020} can not sustain without a certain amount of accretion. In conclusion, TH2 is unlikely to be a  source of UHECRs.

Ultra Luminous X-ray Sources (ULXs), on the other hand,  are promising as UHECR sources, such as NGC 0253 X-1, if we take into account wakefield acceleration. NGC 0253 harbors at least three ULXs with the luminosity ranging between $(2.4-4.1)\times 10^{39}\, \rm erg\, s^{-1}$ \citep{Barnard2010}. The sum of the luminosities of the ULXs reaches $9.1\times 10^{39}\, \rm erg\, s^{-1}$ as shown in Table \ref{tab:bigtable}. They are considered to be intermediate black holes with masses that range from $10^2 - 10^4\, M_\odot$. In fact, we can estimate the mass to be $6.1\times 10^{2} M_{\odot}$ by substituting $\alpha=0.1$, $L_{\rm rad}=9.1\times 10^{39}\, \rm erg\, s^{-1}$, and $\dot{m}=0.1$ into equation \ref{L_rad}. 
Substituting $m=6.1\times 10^2$ and $L_{\rm rad}=9.1\times 10^{39}\, \rm erg\, s^{-1}$, and $\alpha=0.1$ into equations \ref{W_max}, \ref{2pi/omega}, \ref{1/nu}, and \ref{L_gamma}, we can derive $W_{\rm max}=1.8\times 10^{21}\, \rm eV$, $2\pi/\omega=1.5\times 10^{-2}\, \rm s$, $1/\nu=1.4\times10^{-1}\, \rm s$, and $L_\gamma=7.9\times10^{38}\, \rm erg\, s^{-1}$, as shown in Table \ref{tab:bigtable}.

The maximum energy of protons are estimated  to be $W_{\rm max}=1.8\times 10^{21}\, \rm eV$ even less\\ massive black holes ($\sim 600 M_{\odot}$) can generate UHECRs by WFA. The expected UHECR flux ($0.013\,\rm UHECRs/100\,km^{2}/yr$) is consistent with the observed flux ($0.0026\,\rm UHECRs/100\,km^{2}/yr$) for the isotropic distribution, as seen in Table \ref{tab:bigtable} within of a factor of five. 

 The episodic recurrence time is estimated by the WFA theory to be $1/\nu=1.4\times10^{-1}\, \rm s$. \cite{Barnard2010} reports significant variabilities can be seen in 100 second bin, which are much longer when compared with the theoretical predictions. Since observations for very short time variabilities (less than seconds) have unfortunately not been done for the ULXSs in NGC 253, the theory is not constrained by the observations.
 
The theoretical gamma-ray luminosity is estimated as $
L_{\gamma}=7.9\times10^{38}\, \rm erg\, s^{-1}$. The observed gamma ray luminosity (isotropic) is $6.0\times 10^{39}\, \rm erg\, s^{-1}$ in 1-100 GeV \citep{Ackermannetal2012}, which is one order of magnitude higher than the expected gamma-ray flux of ULXs by the wakefield acceleration theory. Direct comparison of $L_\gamma$ luminosity with theory is difficult, however, due to other contributions from other supernova remnants \citep{EichmannandTjus2016}. Furthermore, the jets of NGC 0253 are at a large angle to our line of sight \citep{Aab2018}, and so we see much less luminosity.

\citet{Ebisuzakietal2001} suggested a formation scenario of the central supermassive black holes. 
In this scenario, the ultra-luminous X-ray (ULX) sources considered to be intermediate black holes (IMBH) of the starburst galaxy can eventually collide due to dynamical friction and merge to form a central supermassive black hole.

\section{\label{sec:leve6} Microquasar: SS 433}

SS 433 is a galactic  binary system consisting of a supergiant star $M=10-30 M_\odot$ and a compact object of $M=2-3 M_\odot$ (commonly considered to be a black hole) in the constellation Aquarius. The distance to the S433 system was estimated as 3.5 kpc \citep{BlundelandBowler2004} and is located inside of the supernova remnant W50, which exploded 17-24 thousands years ago \citep{Goodalletal2011}.
SS433 emits jets that have an approximate length of 40 pc, and a bulk velocity of $0.26c$ \citep{Margonetal1984,Fabrika2004}. The two precessing jets model is well established \citep{FabianandRees1979,Milgrom1979,AbellandMargon1979,Katzetal1982,HjellmingandJohnston1981}.  

\citet{Kubotaetal2010} determined the mass of the compact object from orbital analyses to be 2.5 $M_\odot$.  According to \cite{abeysekara18, cherepashchuk05}, the jet luminosity is as high as $10^{40}\, \rm erg\, s^{-1}$, because of super-critical accretion, in spite of very low luminosity ($10^{35-36}\, \rm erg\, s^{-1}$) in X-ray band \citep{Safi-HarbandOegelman1997}. 
Substituting $m=2.5$, $L_{\rm rad}=1.0\times 10^{40}\, \rm erg\, s^{-1}$, and $\alpha=0.1$ into equations \ref{W_max}, \ref{2pi/omega}, \ref{1/nu}, and \ref{L_gamma}, we derived $W_{\rm max}=8.1\times 10^{22}\, \rm eV$, $2\pi/\omega=1.7\times 10^{-2}\, \rm s$, $1/\nu=5.8\times 10^{-4}\, \rm s$, and $L_\gamma=8.7\times10^{38}\,\rm erg s^{-1}$, as shown in Table \ref{tab:bigtable}.

According to the WFA theory, SS433 is capable of accelerating protons, and thus UHECRs; in fact, the maximum acceleration energy $W_{\rm max}$ is as high as $3.5\times 10^{21}\, \rm eV$ (Table \ref{tab:bigtable}). The UHECRs produced in SS433, may not be very localized unfortunately, since it is located near the galactic center, where the magnetic field is higher compared with the outer region. It may produce a broad (more than several ten degrees) concentration toward the galactic center, together with other microquasars in the galactic center region \citep{TYE2020}.

We can also estimate the theoretical recurrence time as $1/\nu=5.78\times 10^{-4}\, \rm s$ according to wakefield theory (Table \ref{tab:bigtable}). Although  a significant variation in flux at the time scale of 10 s was observed \citep{Revnivtsevetal2006}, there is no information in the millisecond range.

The theoretical gamma-ray luminosity is calculated as:
$L_\gamma=8.7\times10^{38}$ erg s$^{-1}$ (Table \ref{tab:bigtable}). Since gamma rays are strongly beamed in the direction of the jets, they are not necessarily seen from Earth; our line of site is not aligned with the jets. The angle between our line of site and the axis of the jet precession is about 74 degree \citep{Davydovetal2008} and the precession angle is about 20 degree \citep{cherepashchuk05}.
SS 433 has been observed to emit gamma rays. First, careful analysis of data from the Fermi gamma-ray observatory Large Area Telescope reveals that SS433 system emits gamma rays with a peak around 250 MeV. It showed a modulation of $\sim 10^{-10}\, \rm erg\, cm^{-2}\,s^{-1}$ correlation with the precession  period  \citep{Rasuletal2019}. The corresponding isotropic luminosity is $3.6\times 10^{37}\, \rm erg\, s^{-1}$. This component may be related to the gamma rays emitted from the electrons accelerated by wakefield in the jets, as suggested by \citet{TYE2020}, though the observed flux is much less compared with the theoretical prediction for the case of isotropic emission.

At $20 \;\rm TeV$, the HAWC detector in \cite{abeysekara18} reported the emission is spatially localized in the three lobes (e1, e2, and w1), 40 pc away from the SS433 system \citep{abeysekara18}. Since the lobes are located where jets interact with the nebula gas, the gamma rays can be explained by the synchrotron emission from high energy electrons accelerated in the wakefiled in the jets, colliding with the magnetic field of $\sim 10\, \mu \rm Gauss$ produced by the interaction of jets with nebula clouds.

Galactic blackhole binaries, such as SS433, Cyg X-1, Cyg X-3, Sco X-1 exhibit relativistic jets,  violent variabilities in time scales ranging from milliseconds to years, and emit radiation from radio to very high energy gamma rays ($\sim \rm TeV$). Because of such non-thermal phenomena, they are considered counterparts of quasars ($\sim 10^{6-9} M_{\odot}$) in million times smaller scales with masses of $\sim 10 M_{\odot}$, in other words, micorquasars, and yet we find they are capable of generating high energy gamma rays and UHECRs.

\section{\label{sec:summary} Summary}
The objects that we have detailed above, in reality, are only a few “good candidates” out of tens and maybe even hundreds of blazars, quasars and microquasars that exhibit WFA in their parsec to kpc scale jets. WFA is almost certainly present in all astrophysical objects that have “jets,” regardless if they are as small as a binary star (10-100 Ms) or as large as blazars ($10^8$ Ms); the parameter values change, but the physics is the same (or similar).

Tables \ref{tab:bigtable} summarize the six specific astrophysical objects in terms of their observed properties along with the theoretically derived values. 
It is still widely believed in the physics and astronomy community that Fermi acceleration is responsible for all high energy cosmic rays, gamma rays, and possibly even UHECRs and UHE gamma rays. We suggest that enough evidence is provided here, and it has been shown before, that it may be too difficult for stochastic Fermi acceleration to explain the creation of UHECRs $> 10^{18}$ eV, and UHE gamma rays $> 10$ GeV in these astrophysical objects, but that WFA can.  Many of these (certainly the six examples examined here) show evidence of being pinpointed origins of high energy gamma rays and UHECRs (and sometimes neutrinos). The Fermi acceleration likely explains less energetic signals such as low energy gamma rays down to radio emissions, before synchrotron radiation becomes insurmountable. 

\begin{figure}[H]
\epsscale{1.0}
\plotone{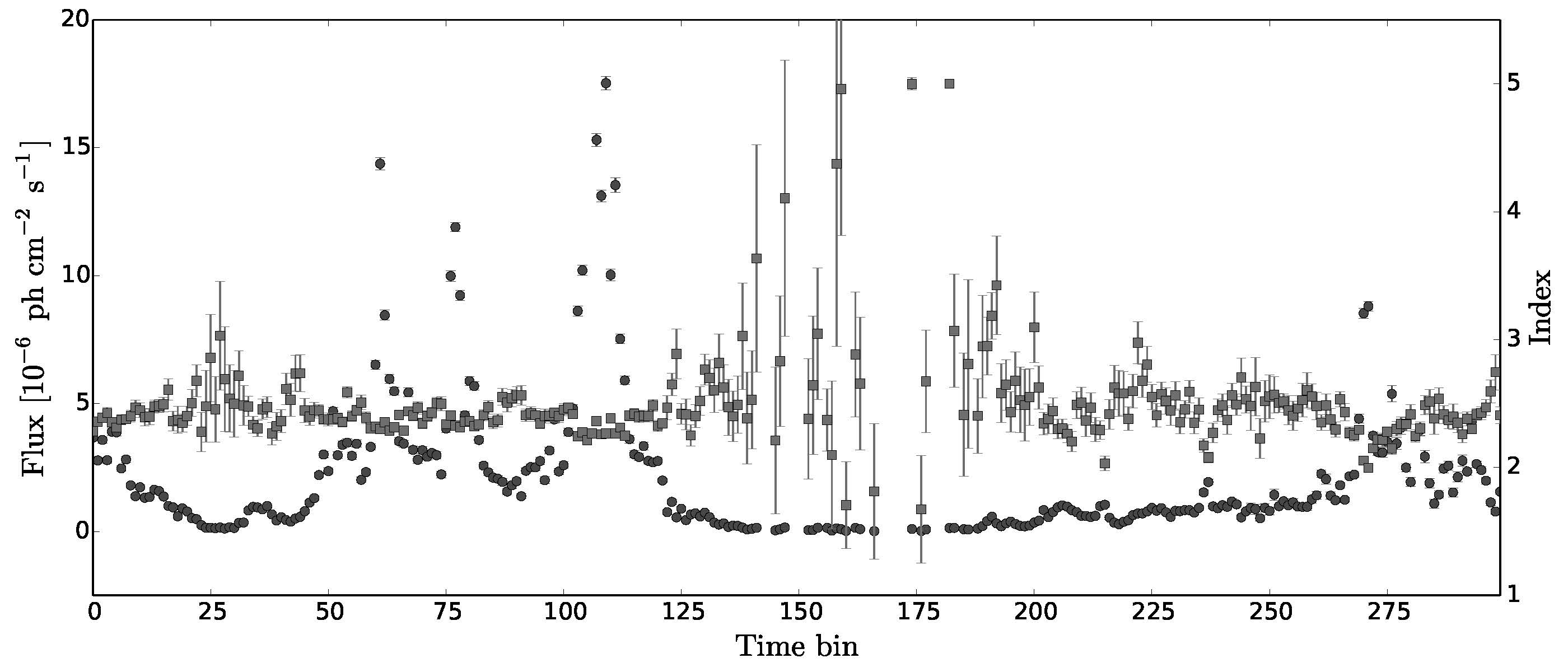}
\caption{\label{fig:flux_index} Flux (circles, left axis) and spectral index (squares, right axis) for 3C 454.3 in 300 time bins of 7.9
days duration. An anti-correlation can be seen: the peaks in flux correspond to dips in the spectral index and vice versa \citep{Canacetal2020}.}
\end{figure}

A linear accelerating mechanism like WFA encounters no such difficulty.  Furthermore, in special cases such as blazars, our theory can even explain time signatures, specifically anti-correlations in the spectral index (\citet{Canacetal2020}; Figure \ref{fig:flux_index}) observed in blazar 3C 453.4 and the flux that Fermi acceleration can say nothing about due to its stochastic, steady state nature. Additionally, neutrino bursts, coincidental with gamma ray bursts, have recently been detected with blazars as the source. Again, to our knowledge only WFA could explain near simultaneous bursts in UHE gamma rays and high energy neutrinos. For objects with distinct knots in their jets, such as Cen A, astronomers have known for decades that the knots within the jets propagate close to the speed of light. This may be understood from WFA that it is natural to have extremely energetic structures that occur in the jets, a part of which contain bow wakes \citep{EbisuzakiandTajima2014a,EbisuzakiandTajima2014b,TYE2020}, which are bound to form propagating dense pockets of electrons (followed by protons) ahead of the pulse.

WFA sheds a new light on interesting time evolving processes, such as fluctuations in the spectra of blazars, and the movement, and acceleration of matter inside astrophysical jets, and now even neutrino bursts. WFA coupled with the magneto-rotational instability (MRI) may be able to provide a virtually complete picture of the generation of UHECRs, UHE gamma rays, and dynamical time signature bursts and fluctuations from start to finish; beginning with MRI causing disk eruption and massive accretion of matter, and ending with the extremely fast particles colliding with decelerated particles in gaseous lobes at the end of the jets, or in Earth's atmosphere as UHECRs.

Though we primarily focus on extragalactic jets as candidates for WFA activities that result in pinpointed emission of UHECRs and correlated (and pronounced structure and localized) emission of gamma rays through secs. 2-7, we also find that jets from much smaller objects, microquasars (such as in our Galaxy) are also capable of emitting UHECRs (including neutrinos) and simultaneously gamma rays. Our theory in fact anticipates more of such microquasars from the Milky Way Galaxy as possible sources of a variety of signals including UHECRs (it would not be easy for Fermi acceleration to take place within the Milky Way Galaxy). 

We note that the spectral index from WFA is also close to 2 (\citep{Mimaetal1991}; similar to that from Fermi mechanism), though we find the spectral index could  vary, such as greater than 2 (see Figure. \ref{fig:flux_index} and  \citet{Canacetal2020}). This means that it is likely that most of the observed (or to be observed) events of UHECRs beyond $10^{18} \rm eV$ arise from WFA, while the events less than this energy can come from both the Fermi mechanism and WFA, as the spectral index alone cannot dissect these events.

We should also note that lower energy phenomena of these objects such as localized emissions of radio waves, soft x-rays with high intensities may be attributable to the disturbances arising not from the jets, but from the accretion disk itself.  This may become natural to understand from the MRI-triggered acceleration (not as well-organized as the wakefield acceleration in the jets we have focused on so far) that arises from the accretion disk itself.  This is briefly explained in figure \ref{fig:MRI}. 
See \citet{Haswell,Okudaetal1992,Smithetal2006, Mineshige1993}.
The time correlations (or lack of them) in this lower energy range of emissions, along with high energy gamma rays and UHECRs (or neutrinos) may provide the subtle relation between the halo dynamics, exterior disk and interior disk, and jets.  In the Appendix we explore a variation introduced by the spread of the jet, which changes theoretical values in Table \ref{tab:bigtable}.

We strongly encourage further verification of wakefield acceleration theory in the universe be conducted via multi-messenger observations including TeV gamma-ray telescope facility, such as the Cerenkov Telescope Array: CTA \citep{CTA2011}, gravitational wave detectors like Advanced LIGO \citep{AdvancedLIGO2015}, the VIRGO \citep{Accadiaetal2011}, and KAGRA  \citep{Akutsuetal2018}, and  ultra high energy cosmic rays/neutrino observatories, such as Probe of Extreme Multi-Messenger Astrophysics: POEMMA \citep{Anchordoqui2020}, and The Giant Radio Array for Neutrino Detection (GRAND) \citep{Alvarez-Muniz2019}. We hope WFA theory and the work presented here contribute to these observational efforts. 

\acknowledgments

We would like to thank Professors  S. Bird, S. Barwick, K. Abazajian, H. Sobel, J. Bullock, G. Mourou, X.Q. Yan, T. Tait, S. Murgia, P. Picozza, P. Klimov, S. Bulanov, T. Esirkepov, K. Nakajima, the late Y. Takahashi, the late J. A. Wheeler, and the JEM-EUSO collaboration for their kind discussions that helped carry out this work. This work started as a term project report by students who participated in the University of California, Irvine graduate course Physics 249 ``Plasma Astrophysics" (Spring, 2020), in which the instructor (TT) guided the subject of wakefields in astrophysics. This work was supported by the Rostoker Fund.

\bibliography{Huxtableetal2020.bib}
\bibliographystyle{aasjournal}



\appendix
\section{\label{sec:leve7} Dependence of Wakefield on the Jet Spread}

As we have examined in previous sections \ref{sec:Blazar}-\ref{sec:leve6}, in interpreting the individual astrophysical objects and phenomena only one model may serve as sufficient in understanding those astrophysical objects. While the general theory we described through the excitation of the disk disturbances, including shaking of the jets, and the subsequent wakefield generation and acceleration of particles along the jets have turned out to be quite generic and deep rooted physics common among these objects and their phenomena, despite its disparate scales and mass differences of the central objects.  There are some important individualities that may matter in detailed manifestations of the objects and parameters. One example of such may be the jet’s spreading angle. Jets may be strongly collimated by the spiraling surrounding magnetic fields. This may relax certain constraints on phenomena and parameters. Here we introduce the jet spreading by one model parameter of the power index that determines the jet diameter as a function of the distance from the central object, as the jet particles and magnetic fields emanate outward.


We discuss the dependence of physical parameters in the jet on distance from the
bottom and discusses how the waves propagate through it. First, we assume that
\def\theequation{A\arabic{equation}}
\begin{equation}
    b(D)=R_0m(D/R_0m)^p
\end{equation}
Although the power law index $p$ is observed to be close to $\sim 0.5$ for the case of M87, the closest active galactic nuclei M87 \citep{AsadaandNakamure2012} and many other AGN jets \citep{Pushkarevetal2017}, it may be different in the range of 0 (a cylinder) to 1 (a linear cone).

The cyclotron frequency $\omega_c'$ in the jet corrected for relativistic effects is given by

\begin{equation}
    \omega_{\rm c}'= \frac{eB_{\rm jet}}{m_{\rm e}c\gamma}
\end{equation}

On the other hand, the magnetic field $B_{\rm jet}$ in the jet can be calculated assuming that the magnetic field flux is conserved in the jet.

\begin{align}
    B_{\rm jet} &=[B_{\rm disk}(r=1)](b/mR_0)^{-2}\\
    &=[B_{\rm disk}(r=1)]\bigg(\frac{D}{mR_0}\bigg)^{-2p}\\
    &= \bigg( \frac{16\pi c^2}{3\sqrt{6}\kappa_{\rm T}R_0}\bigg)^{1/2}m^{-1/2}\bigg( \frac{D}{mR_0}\bigg)^{-2p}
\end{align}

Next, we assume as
\begin{equation}
    \gamma = a_0
\end{equation}

within the jet, $a_0$ can be calculated, assuming that the wave intensity within the jet is conserved, {\it i.e.}, the flux $\phi_{\rm w,jet}$ is inversely proportional to the cross-sectional area $\pi b^2$ of the jet.
\begin{equation}
    a_0(D)=a_0(D=R_0)\bigg( \frac{b(D)}{R_0m}\bigg)^{-1}
\end{equation}

where $D$ is the distance from the bottom of the jet, and $b(D)$ is the radius of the jet, which is assumed to be $b(0)=3R_g=R_0m$. In addition, Figure 2 shows the ratio $\omega_c'/\omega$ of the cyclotron frequency to the wave frequency and that of plasma frequency $\omega_p'/\omega$, plotted against the distance $D/(R_0m)$ from the bottom of the jet for the typical cases ($\Gamma=10$, $\alpha=0.1$, $\xi=10^{-2}$, $\dot{m}=0.1$, and $m=1,10^4,10^8$). Here Now, we get
\begin{equation}
    a_0(D)=\frac{\rm e}{36m_{\rm e}c}\sqrt{\frac{R_0}{\pi\epsilon^3\kappa_{\rm T}}}\alpha^{3/4}
    \dot{m}^{3/2}m^{1/2}
    \bigg( \frac{D}{R_0m}\bigg)^{-p}
\end{equation}

Substituting equations 3, 4, and 6 into equation 2, we obtain
\begin{equation}
    \omega_{\rm c}'=\frac{144c\pi}{R_0}\bigg( \frac{\epsilon^3}{3\sqrt{6}}\bigg)^{1/2}
    \frac{1}{\alpha^{3/4}\dot{m}^{3/2}m}\bigg( \frac{D}{R_0m}\bigg)^{-2p}
\end{equation}

On the other hand, the plasma frequency $\omega_{\rm p}'$ corrected for relativistic effects is given by
\begin{equation}
    \omega_{\rm p}'= \bigg( \frac{4\pi n_{\rm jet}e^2}{m_{\rm e}\gamma\Gamma^3}\bigg)^{1/2}
\end{equation}

The plasma density $n_{\rm jet}$ in the jet can be calculated from as follows, if we assume the kinetic luminosity of the jet:
\begin{equation}
    L_{\rm jet}= n_{\rm jet}\mu m_{\rm H}c^3\Gamma^2\pi b^2 = \xi L_{\rm rad}
\end{equation}

is conserved through the jet.
\begin{equation}
    n_{\rm jet} = \frac{2}{3 \mu m_{\rm H}\kappa_{\rm T} R_0}\frac{\xi \dot{m}}{\Gamma^2 m}\bigg( \frac{D}{R_0m}\bigg)^{(p-1)/2}
\end{equation}

Here, $\xi$ is the ratio of the kinetic luminosity of the jet to the radiation luminosity, $\Gamma$ is the bulk Lorentz factor, and $\mu = 1.29$ is the mean molecular weight of the accreting gas. Substituting equations 10, 4, and 6 into equation 8, we get:
\begin{align}
    \omega_{\rm p}' &= \bigg( \frac{4\pi n_{\rm jet}e^2}{m_{\rm e}\gamma\Gamma^3}\bigg)^{1/2}\\
    &= \bigg( \frac{96\pi ec}{\mu m_{\rm H}}\bigg)^{1/2}\bigg( \frac{\pi\epsilon^3}{R_0\kappa_{\rm T}}\bigg)^{1/4}
    \frac{\xi^{1/2}}{\Gamma^{5/2}\alpha^{3/8}\dot{m}^{1/4}m^{3/4}}
    \bigg( \frac{D}{R_0m}\bigg)^{(p-1)/2}
\end{align}

 For most of the interesting cases, the relationship of $\omega_{\rm c}'$, $\omega_{\rm p}'>\omega$ holds; In other words, at the bottom of the jets, the plasma in the over dense state ($\omega_{\rm p}'>\omega$), where plasma waves and electromagnetic waves cannot propagate. On the other hand, Alfvén wave or whistler wave cannot propagate, since $\omega'_{\rm c}>\omega$, the Alfvén velocity $V_{a,jet}$ at the bottom of the jet are given by
\begin{equation}
    V_{\rm A,jet}=\frac{B_{\rm jet}}{\sqrt{4\pi m_{\rm H}n_{\rm jet}}}=\big( \frac{2}{\sqrt{6}}\big)^{1/2}
    c\frac{\Gamma}{\xi^{1/2}\dot{m}^{1/2}}
\end{equation}

In other words, the nominal values of the Alfvén velocity 
\begin{equation}
    V_{\rm A,jet}\sim 10^{12}[\rm cm\, s^{-1}]\bigg( \frac{\Gamma}{10}\bigg)\bigg( \frac{\xi}{10^{-2}}\bigg)^{-1/2}
\end{equation}

This can approach the speed of light, when the approximation breaks down. Then the wave becomes that of EM waves in magnetized plasma.  On the other hand, $\omega_{\rm p}'=\omega$ at the distance $D_2$ given by:
\begin{equation}
    \bigg( \frac{D_2}{R_0m}\bigg) = \bigg[ \frac{4R_0e^2}{9\pi\mu^2m_{\rm H}^2c^2\epsilon}
    \frac{\xi^2\alpha^{1/2}\dot{m}^3m}{\Gamma^{10}}\bigg]^{2(1-p)}
\end{equation}

On the outside of the point $D_2\ (D>D_2),\ \omega>\omega_{\rm p}'$ so that the plasma wave (electromagnetic wave) is allowed to propagate. The electromagnetic waves propagated as Alfvén and whistler waves are converted into plasma waves (electromagnetic waves) by nonlinear mode-conversion. This $D>D_2$ leads to the bow wakefield acceleration as described in the next subsection.

The pondermotive force, $F_{\rm pm}$, which acts on the electrons caught in an intense electromagnetic wave is a force generated from the Lorentz force, $\big( \frac{v}{c}\big) \times B$, in the propagation direction of the electromagnetic wave. If the motion of the electrons by the wave is not relativistic ($a<1$), it can be calculated as the force resulting from the average of the profiles of the electromagnetic pulses. In the relativistic regime ($a>1$), this force is more simplified.  Since the particle velocity asymptotically approaches the light velocity and if the plasma satisfies the under dense ($\omega > \omega_p'$)
condition as well, then $B=E$. In this case, $F_{\rm pm}$, is given by
\begin{equation}
    F_{\rm pm} = \Gamma m_{\rm e}eca\omega
\end{equation}

Charged particles are accelerated by an electric field generated by bow wakefield (longitudinal polarization of electronic distributions). As shown in Fig. \ref{fig:bowwake}, protons are accelerated at the back slope of the wakefield, while electrons are accelerated at the front slope. The acceleration force $F_{\rm acc}$ is given by
\begin{align}
    F_{\rm acc} &= zF_{\rm pm} = z\Gamma eE_{\rm w}\bigg( \frac{D}{R_0}\bigg)^{-p}\\
    &= \frac{ec}{3}\bigg( \frac{\pi}{\epsilon\kappa_{\rm T} R_0}\bigg)^{1/2}
    \frac{z\Gamma\alpha^{1/4}\dot{m}^{1/2}}{m^{1/2}}\bigg( \frac{D}{R_0m}\bigg)^{-p}
\end{align}

Here $z$ is the charge of the particle.  The maximum energy, $W_{\rm max}$, obtained by the particle is determined by integrating $F_{\rm acc}$ over the acceleration distance, $D_3$
\begin{align} \label{Wmax}
    W_{\rm max} &= \int^{D_3}_0 F_{\rm acc}dD \\
    &= \frac{ec}{3}\bigg( \frac{\pi}{\epsilon\kappa_{\rm T} R_0}\bigg)^{1/2}\frac{z\Gamma\alpha^{1/4}\dot{m}^{1/2}}{m^{1/2}}
    \int^{D_3}_0 \bigg( \frac{D}{R_0m}\bigg)^{-p}dD\\
    &= \frac{2ec}{3}\bigg( \frac{\pi R_0}{\epsilon\kappa_{\rm T}}\bigg)^{1/2}
    z\Gamma\alpha^{1/4}\dot{m}^{1/2}m^{1/2}
    \bigg( \frac{D_3}{R_0m}\bigg)^{1-p}
\end{align}

The acceleration distance, $D_3$, is evaluated as: 
\begin{equation} \label{D3eqn}
    D_3 = \frac{e}{432m_{\rm e}c}\bigg( \frac{R_0^3}{\pi^3\epsilon^5\kappa_{\rm T}}\bigg)^{1/2}
    \alpha^{5/4}\dot{m}^{5/2}m^{3/2}\bigg( \frac{D_3}{R_0m}\bigg)^{-p}
\end{equation}

We can solve equation \ref{D3eqn} for $(D_3/R_0m)$
\begin{align} \label{D3eqn2}
    \bigg( \frac{D_3}{R_0m}\bigg) =  \bigg(\frac{e}{432m_{\rm e}c}\bigg)^{1/(1+p)}\bigg(\frac{R_0}{\pi^3\epsilon^5\kappa_{\rm T}}\bigg)^{1/2(1+p)}  \alpha^{5/4(1+p)}\dot{m}^{5/2(1+p)}m^{1/2(1+p)}
\end{align}

Substituting equation \ref{D3eqn2} into equation \ref{Wmax}, we obtain
\begin{equation}
    W_{\rm max} = \frac{1}{3^{(4-2p)/(1+p)}}
    \bigg( \frac{\pi^{-2+4p}e^4c^{4p}R_0^2}{2^{2-2p}m_{\rm e}^{2(1-p)}\epsilon^{6-4p}\kappa_{\rm T}^2}\bigg)^{1/2(1+p)} z\Gamma\alpha^{(6-4p)/4(1+p)}\dot{m}^{(6-4p)/2(1+p)}m^{1/(1+p)}
\end{equation}

Here we can eliminate $\dot{m}$ as
\begin{align}
    W_{\rm max}=\frac{1}{6}\bigg( \frac{3^{2p}e^4\kappa_{\rm T}^{4-4p}}{4\pi^{8-8p}m_{\rm e}c^{18-16p}R_0^{4-4p}\epsilon^{6-4p}}\bigg)^{1/2(1+p)}
     z\Gamma\alpha^{(6-4p)/4(1+p)}m^{-2(1-p)/(1+p)}L_{rad}^{(6-4p)/2(1+p)}
\end{align}

The p-dependent formulae in this section can be naturally reduced, if we take $p=1/2$, in the ``standard case" for example, like in the previous sections.  It may be useful to incorporate such a dependency on the jet spreading ({\it p}-index) in assessing maximum proton energy from SS433, as well as other astrophysical objects.




\end{document}